\documentclass[pdflatex,sn-mathphys-ay]{sn-jnl}% Math and Physical Sciences Numbered Reference Style

\usepackage{multirow}
\usepackage{xcolor}
\usepackage{natbib}
\usepackage{graphicx}%
\usepackage{multirow}%
\usepackage{amsmath,amssymb,amsfonts}%
\usepackage{amsthm}%
\usepackage{mathrsfs}%
\usepackage[title]{appendix}%
\usepackage{xcolor}%
\usepackage{textcomp}%
\usepackage{manyfoot}%
\usepackage{booktabs}%
\usepackage{algorithm}%
\usepackage{algorithmicx}%
\usepackage{algpseudocode}
\usepackage{listings}%
\usepackage{caption}
\usepackage{subcaption}
\makeatletter
\renewcommand{\bmhead}[1]{{\bfseries #1}}
\makeatother
\usepackage[british]{babel}
\usepackage{microtype}
\usepackage[none]{hyphenat}
\usepackage[dvipsnames]{xcolor}
\usepackage[commandnameprefix=always, final]{changes}
\robustify\cite
\begin{document}

\title[Article Title]{Type II and Type III Solar Radio Burst Classification Using Transfer Learning}

%%=============================================================%%
%% GivenName    -> \fnm{Joergen W.}
%% Particle    -> \spfx{van der} -> surname prefix
%% FamilyName    -> \sur{Ploeg}
%% Suffix    -> \sfx{IV}

\author*[1,2,6]{\fnm{Herman} \sur{le Roux}}\email{herman.leroux@dias.ie}

\author[3]{\fnm{Ruhann} \sur{Steyn}}\email{Ruhann.Steyn@nwu.ac.za}

\author[3,4]{\fnm{Du Toit} \sur{Strauss}}\email{Dutoit.Strauss@nwu.ac.za}

\author[1]{\fnm{Mark} \sur{Daly}}\email{Mark.Daly@tus.ie}

\author[2]{\fnm{Peter T.} \sur{Gallagher}}\email{peter.gallagher@dias.ie}

\author[1]{\fnm{Jeremiah} \sur{Scully}}\email{Jeremiah.Scully@tus.ie}

\author[2]{\fnm{Shane A.} \sur{Maloney}}\email{shane.maloney@dias.ie}

\author[5]{\fnm{Christian} \sur{Monstein}}\email{cmonstein@swissonline.ch}

\author[6]{\fnm{G{\"u}nther} \sur{Drevin}}\email{Gunther.Drevin@nwu.ac.za}
\equalcont{These authors contributed equally to this work.}

\affil[1]{\orgname{Faculty of Engineering and Informatics, Technological University of the Shannon}, \city{Athlone}, \country{Ireland}}
\affil[2]{\orgname{Astronomy and Astrophysics Section, School of Cosmic Physics, Dublin Institute for Advanced Studies, DIAS Dunsink
Observatory}, \city{Dublin}, \country{Ireland}}

\affil[3]{\orgname{Center for Space Research, Faculty of Natural and Agricultural Sciences, North-West University}, \city{Potchefstroom}, \country{South Africa}}

\affil[4]{\orgname{National Institute for Theoretical and Computational Sciences (NITheCS)}, \city{Potchefstroom}, \country{South Africa}}

\affil[5]{\orgname{Istituto ricerche solari Aldo e Cele Daccò (IRSOL), Università della Svizzera italiana}, \city{Locarno}, \country{Switzerland}}

\affil[6]{\orgname{School of Computer Science and Information Systems, Faculty of Natural and Agricultural Sciences, North-West University}, \city{Potchefstroom}, \country{South Africa}}

\abstract{The Sun periodically emits intense bursts of radio emission known as solar radio bursts (SRBs). These bursts can disrupt radio communications and be indicative of large solar events that can disrupt technological infrastructure on Earth and in space. The risks posed by these events highlight the need for automated SRB classification, providing the potential to improve event detection and real-time monitoring. This would advance the techniques used to study space weather and related phenomena. A dataset containing images of radio spectra was created using data recorded by the Compound Astronomical Low frequency Low cost Instrument for Spectroscopy and Transportable Observatory (e-Callisto) network. This dataset comprises three categories: empty spectrograms; spectrograms containing Type II SRBs; and spectrograms containing Type III SRBs. These images were used to fine-tune several popular pre-trained deep learning models for classifying Type II and Type III SRBs. The evaluated models included VGGnet-19, MobileNet, ResNet-152, DenseNet-201, and YOLOv8. Testing the models on the test set produced F1 scores ranging from 87\% to 92\%. YOLOv8 emerged as the best-performing model among them, demonstrating that using pre-trained models for event classification can provide an automated solution for SRB classification. This approach provides a practical solution to the limited number of data samples available for Type II SRBs.}

\keywords{solar radio bursts; convolutional neural networks; transfer learning.}

\maketitle

\section{Introduction}\label{sec1}

\noindent Space weather refers to the impact of solar activity, such as particle emissions caused by solar flares and coronal mass ejections (CMEs), when these phenomena travel through space and interact with the solar system and our modern technological infrastructure \citep{temmer2021space,li2024predicting,liang2024predicting}. During these events, electrons are accelerated in the solar corona and generate radio emissions called solar radio bursts (SRBs) as they propagate away from the Sun \citep{1985ARA&A..23..169D,pick2008sixty,morosan2023type,gary2023new}.\\

\begin{figure}[h]
  \centering
  % Top row: Original spectrograms
  \begin{subfigure}{0.3\linewidth}
    \centering
    \captionsetup{position=top, justification=raggedright}
    \caption{Empty}
    \includegraphics[width=\linewidth, keepaspectratio]{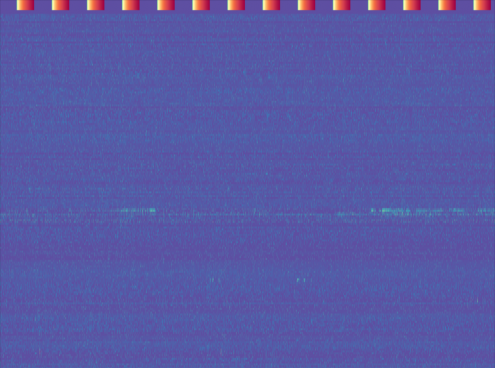} 
    \label{fig:TypeIIEgypt}
  \end{subfigure}\hfill
  \begin{subfigure}{0.3\linewidth}
    \centering
    \captionsetup{position=top}
    \caption{Type II}
    \includegraphics[width=\linewidth, keepaspectratio]{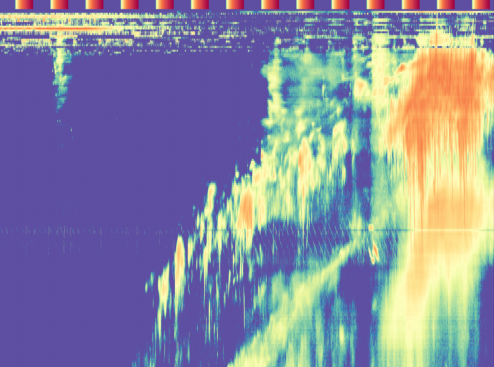}
    \label{fig:TypeIIAus}
  \end{subfigure}\hfill
  \begin{subfigure}{0.3\linewidth}
    \centering
    \captionsetup{position=top}
    \caption{Type III}
    \includegraphics[width=\linewidth, keepaspectratio]{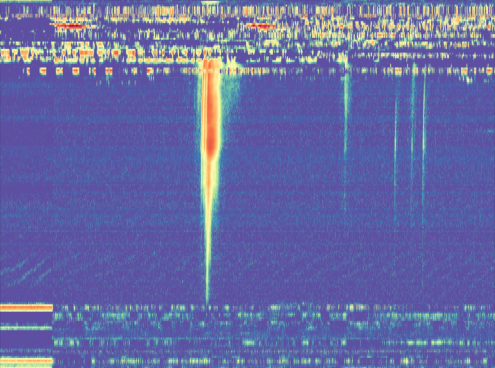}
    \label{fig:TypeIIIBir}
  \end{subfigure}
  \vspace{1em} % Space between rows
  % Bottom row: Processed spectrograms
  \begin{subfigure}{0.3\linewidth}
    \centering
    \captionsetup{position=top}
    \caption{Processed Empty}
    \includegraphics[width=.8\linewidth, keepaspectratio]{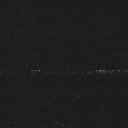}
    \label{fig:TypeIIEgyptgrey}
  \end{subfigure}\hfill
  \begin{subfigure}{0.3\linewidth}
    \centering
    \captionsetup{position=top}
    \caption{Processed Type II}
    \includegraphics[width=.8\linewidth, keepaspectratio]{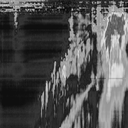}
    \label{fig:TypeIIAusgrey}
  \end{subfigure}\hfill
  \begin{subfigure}{0.3\linewidth}
    \centering
    \captionsetup{position=top}
    \caption{Processed Type III}
    \includegraphics[width=.8\linewidth, keepaspectratio]{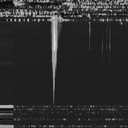}
    \label{fig:TypeIIIBirgrey}
  \end{subfigure}  \caption{Spectrograms and their processed versions depicting Empty, Type II, and Type III solar radio burst classes. (a) Empty sample observed at the e-Callisto spectrometer in Alexandria, Egypt. (b) Type II solar radio burst observed in Sunnydale, Australia. (c) Type III solar radio burst observed in Birr, Ireland. (d) Processed spectra of an Empty sample from Alexandria, Egypt. (e) Processed spectra of a Type II solar radio burst from Sunnydale, Australia. (f) Processed spectra of a Type III solar radio burst from Birr, Ireland.}
  \label{fig:SolarBurstsCombined}
\end{figure}

\noindent SRBs are classified into five main types, labelled from I to V,  according to the unique characteristics \citep{wild1963}. This study focuses primarily on Types II and III SRBs. Type II SRBs typically drift from high to low frequencies at a slower rate (see Figure \ref{fig:TypeIIAus}) and are known as `slow-drifting bursts' \citep{wild1963}. Although they can occur without the presence of an apparent CME (\citealt{2009Gopalswamy,feng2025multiple}), Type II bursts are often linked to shock waves generated by CMEs (\citealt{morosan2023type,kumari2025type}). In spectrograms, Type II bursts appear as sloped, slow-drifting flux.\\

\noindent  In contrast, Type III SRBs drift from high to low frequencies at a much higher rate (see Figure \ref{fig:TypeIIIBir}) and appear as rapidly drifting streaks \citep{zucca2012observations}. These Type III SRBs can also appear in groups that, when dense enough, can resemble the structure of Type II bursts. Type III bursts are usually linked to energetic electrons ejected during solar flares and travelling along open or quasi-open magnetic field lines \citep{cane1984,Reid2014,2021Chen,chen2024spectral,feng2024solar}. Their strong correlation with such intense solar activities makes these bursts particularly valuable for researching space weather phenomena.\\ 

\noindent These SRBs are currently classified manually, a laborious and error-prone process given the volume and complexity of the data. An automated approach could produce better results and reduce the time required to categorise SRBs, thereby improving space weather event research. One such approach is to use machine learning (ML) methods. \\

\noindent ML is the development of models that can recognise patterns in data and use that information to independently make predictions or forecasts on new and unseen data \citep{kufel2023machine}. ML is usually divided into different learning types, such as supervised, unsupervised, and reinforcement learning, depending on the task \citep{aima}. Variants such as semi-supervised, self-supervised, and transfer learning further refine this framework, with supervised and transfer learning relevant to SRB classification in this study. Deep learning is a specialised subset of ML models that utilises deep neural networks. \\

\noindent A supervised learning model uses a dataset of input-output pairs to learn a function that maps inputs to the correct outputs. In this context, each output (or ground truth) is typically human-labelled, with an unknown function underlying the relationship. The effectiveness of this function depends on its ability to generalise well to new and unseen data, a primary goal in supervised learning \citep{chollet2021}.\\

\begin{figure}[h]
    \centering
    \includegraphics[width=1\textwidth]{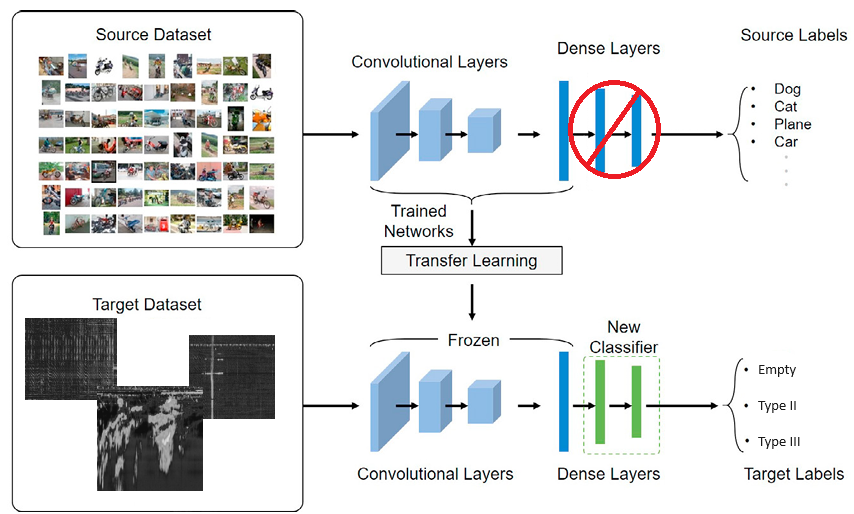}
    \captionsetup{width=0.9\textwidth}
    \caption{A diagram of the general architecture of transfer learning models, illustrating the frozen feature extraction layers and retrained dense layers, which have been adapted for this study \citep{tao2020real}.}
    \label{fig:transferlearningdiagram}
\end{figure}

\noindent Transfer learning leverages knowledge from a previously trained model, typically trained using supervised learning, to tackle a different task \citep{weiss2016survey}. In neural networks, this is achieved by freezing most layers of a pre-trained model and fine-tuning a few final layers to create a new target network, as illustrated in Figure \ref{fig:transferlearningdiagram} \citep{zhuang2020comprehensive}. This technique is beneficial when datasets are small or resource constraints limit data collection, as is the case for Type II SRBs \citep{weiss2016survey}. It has also proven effective in tasks such as image classification, leveraging a diverse set of pre-trained Convolutional Neural Network (CNN) architectures \citep{simonyan2014VGG,he2016deepResnet,howard2017mobilenets,huang2017densely}. \\

\noindent As with most modern fields, data-driven techniques such as ML, specifically supervised and transfer learning, offer promising pathways for developing models to classify different types of SRB. Enhancing these classification capabilities could significantly advance research in this field.\\

\noindent The approach in this study uses CNN architectures and transfer learning to automate the classification of SRBs. CNNs are a popular deep learning architecture for computer vision, using kernels to convolve through images and identify features such as edges and objects \citep{o2020deep}. CNNs are typically composed of three primary types of layers. Convolutional layers are the first layers and apply kernels that traverse the input data to produce feature maps \citep{albawi2017understanding}. Pooling layers are introduced to down-sample feature maps \citep{o2020deep}. Finally, fully connected layers, also called dense layers, condense feature maps into feature vectors to learn global patterns and relationships among features \citep{chollet2021}.\\ 

\noindent Recent years have witnessed significant progress in applying ML techniques to SRB analysis, though important distinctions exist between detection and classification tasks. Several studies have applied ML techniques to SRB analysis. Table \ref{tab:related_work} summarises recent work in this area, distinguishing between detection and classification approaches and highlighting how their goals differ from the current study.\\

\begin{table}[h]
    \centering
    \caption{Summary of recent ML approaches for SRB analysis, their performance metrics, and how their objectives differ from this study.}
    \label{tab:related_work}
    \begin{tabular}{@{}p{4cm}p{2cm}p{2.25cm}p{3.5cm}@{}}
\toprule
        \textbf{Study} & \textbf{Approach} & \textbf{Performance} & \textbf{Task} \\
\midrule
        \cite{guo2022deep} & CNN & 98.73\% accuracy & Detection of any SRBs; does not classify by SRB type \\
        \cite{scully2023improved} & Object detection & 77.71\% mAP & Detection and localisation of Type III SRBs only  \\
        \cite{orue2023automaticCURRENTRESEARCH} & AlexNet, VGG16 & 90.7\% accuracy & Detection of Type III SRBs only  \\
        \cite{zhao2023solar} & CNN, MobileNet & 98.95\% accuracy & Classification of Type III and Type IV\\
        \cite{bussons2023automatic} & Detection model & 93\%--96\% accuracy & General burst detection to identify missed events and not focused on type classification \\
        \cite{zhang2024identificationCURRENTRESEARCH} & YOLOv7 & 73.5\% mAP & Detection of Type II and III SRBs\\
        \cite{murphy2024automaticCURRENTRESEARCH} & UNET  & 93\% F1 score & Detection and segmentation of Type III SRBs\\
        \cite{SRBObjectDetecWang,SRBDetecDETRWang} & Object detection models & 79.9\% and 83.5\% mAP@50, 95.1\% and 99.4\% recall & Multi-class detection however 92.5\% of data samples were Type III events. \\
\bottomrule
    \end{tabular}
\end{table}

\noindent Classifying SRBs is faced with limited sampling and imbalanced training data sets. To overcome this, this work utilises a stratified sampling strategy to construct a balanced training dataset from spectrogram images sourced from the Compound Astronomical Low frequency Low cost Instrument for Spectroscopy and Transportable Observatory (e-Callisto;  \url{https://www.e-callisto.org}) network \citep{benz2009world}. We introduce the application of transfer learning to fine-tune established CNN architectures, including VGGnet-19, MobileNet, ResNet-152, DenseNet-201, and YOLOv8. This differs from previous approaches that employ augmentation methods to generate new data samples.  Class-specific performance metrics show that this technique effectively handles the limited data availability for Type II SRBs while simultaneously mitigating the significant noise and observational variability introduced by the e-Callisto network's multiple global stations.\\

\noindent In this paper, we propose transfer learning as a reliable approach for classifying Type II and Type III SRBs despite the limited number of available data samples for Type II SRBs. In section \ref{Sec:methods} we describe the detailed methods, beginning with the collection of data from the publicly available e-Callisto data repository. This section further details the specific pre-processing and labelling techniques used to create the balanced training dataset, along with a brief overview of the e-Callisto network itself. Furthermore, the specific training environment, model architectures, and hyperparameters are detailed. We then present the results in Section \ref{Sec:results}, discussing the performance metrics obtained by fine-tuning various CNN architectures on the prepared dataset, which demonstrates the models' ability to classify the three spectrogram classes. Finally, in Section \ref{Sec:conclusion}, we conclude, summarising the key findings and highlighting potential improvements for future work.\\

\section{Methods}\label{Sec:methods}

\noindent This section outlines the methodology used in this study, including the dataset, model training process, and performance evaluation metrics. The data were collected from publicly available e-Callisto data, pre-processed, and labelled to create a balanced training dataset \citep{monstein2023callisto}. The e-Callisto network is a global network of solar radio spectrometers designed to monitor SRBs and radio frequency interference (natively in the 45–870 MHz range) using low-cost, programmable heterodyne receivers for space weather research and astronomical science \citep{benz2009world}.\\

\noindent All models were developed in Python and trained on a system equipped with an Intel Core i7-12700 processor, 32 GB DDR4 RAM, and an Nvidia RTX 3090 Ti GPU. The software was created using Anaconda for environment management, Visual Studio Code for code editing, and Jupyter Notebooks for interactive development. The models were created using the TensorFlow \citep{tensorflow2015-whitepaper}, Keras \citep{chollet2015keras}, and PyTorch \citep{DBLP:journals/corr/abs-1912-01703} libraries.\\

\noindent Several transfer learning models, including VGGNet-19 \citep{simonyan2014VGG}, MobileNetV2 \citep{he2016deepResnet}, ResNet-152 \citep{howard2017mobilenets}, DenseNet-201 \citep{huang2017densely}, and YOLOv8 \citep{Jocher_Ultralytics_YOLO_2023}, were trained using this dataset. The models were evaluated using key performance metrics, such as accuracy, precision, recall, and F1 score, to assess their effectiveness in classifying SRB events. \\

\subsection{Dataset}

\noindent The dataset for this study was a curated version of a previously created dataset, compiled from publicly available data in the e-Callisto repository \citep{leRoux2022}. The repository's files, stored in Flexible Image Transport System (FITS) format, include date, time, and sensor data. These files were processed into images of events recorded by radio frequency spectrometers (see Figure \ref{fig:SolarBurstsCombined}). The samples were collected at 15-minute intervals and recorded in UTC. These samples are recorded simultaneously at multiple stations worldwide, each with varying installations and noise levels. All of the data used in this study were recorded from 2021-01-01 to 2023-04-21 .\\

\begin{figure}[h]
    \centering
    \includegraphics[width=\textwidth]{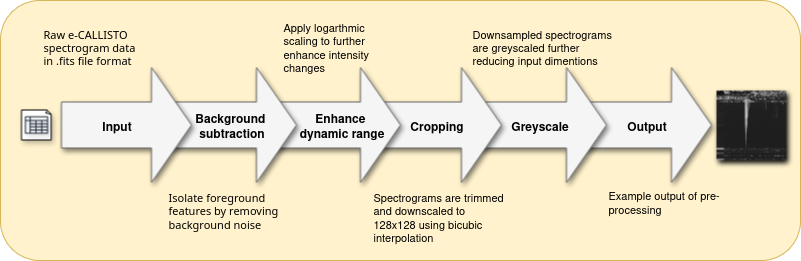}
    \captionsetup{width=0.9\textwidth}
    \caption{Flowchart of the pre-processing steps applied to the data.}
    \label{fig:flowchart}
\end{figure}

\noindent The data samples were pre-processed as described in Figure \ref{fig:flowchart} to produce spectrograms. These images were cropped to include only the data, excluding elements such as titles and intensity-colour scales, resulting in 324x357 pixel images. The images were then resized from 324x357 pixels (99 284 pixels per image) to 128x128 pixels (16 384 pixels per image) using bicubic interpolation, resulting in an 85.84\% reduction in pixel data. The images were converted to greyscale using the Pillow library, storing only intensity values (0-255) in a single channel to simplify the input shape \citep{clark2015pillow}. This pre-processing step was applied to reduce the resources required during training. The dataset collected comprised $\approx$20,500 images, reduced from an initial $\approx$28 000 due to corruption caused by incorrectly stored header data. To maintain consistency for comparison, corrupted samples were removed.\\

\noindent The data samples were labelled using event lists compiled by the e-Callisto principal investigator, Christian Monstein, a former engineer at the Istituto Ricerche Solari Locarno (IRSOL). Events of interest, specifically Type II and Type III SRBs, were isolated and assigned to their respective classes. Additionally, a third class, referring to empty spectrograms, was created, consisting of randomly selected samples without recorded or labelled events of interest.\\

\noindent To evaluate each model's performance, a balanced dataset was created using stratified sampling. This ensured that an over-represented class did not bias the model's performance. The Type II class, having the fewest samples ($\approx$1200), was used as the reference point, and the other classes ($\approx$16,500 Type III samples and $\approx$2800 Empty spectrograms) were randomly downsampled to match their size. The resulting dataset contained 3561 images. From this, 100 samples from each class were randomly selected to form a test set. The remaining samples were divided into training and validation sets using a 4:1 ratio. \\

\subsection{Model training}
\noindent In this experiment, multiple transfer learning architectures were trained using the stratified dataset to compare their performance. All models were pre-trained on the ImageNet dataset, with feature extraction layers frozen and fully connected layers fine-tuned on the newly created dataset \citep{deng2009}.\\ 

\noindent The architectures used included VGGNet-19, MobileNetV2, ResNet-152, DenseNet-201, and YOLOv8. Each model’s base layer was connected to a flattening layer, followed by a fully connected dense layer with 128 nodes, two dropout layers (rate = $0.5$), and a final dense layer with three nodes activated by a softmax function for predictions. Models were compiled with the Adam optimiser function (learning rate = $2\times10^{-4}$) and the categorical cross-entropy loss function.\\

\noindent Each of these architectures employs distinct design principles:
\begin{itemize}
    \item VGGNet-19 uses sequentially stacked convolutional layers with small kernels.
    \item MobileNetV2 prioritises lightweight design for resource-limited environments, employing depthwise separable convolutions to reduce computational cost.
    \item ResNet-152 incorporates skip connections to learn residual mappings, enhancing training efficiency and mitigating vanishing gradient issues.
    \item DenseNet-201 uses densely connected blocks, enabling feature reuse and improved gradient flow through direct connections between layers within a block.
\end{itemize}

\noindent The final model was trained using the YOLOv8 architecture. The YOLOv8 architecture was also trained using the stratified dataset. The YOLOv8 architecture comprises the backbone, neck, and head. These components are used to extract features from the data, reduce the spatial dimensions, and finally pass the data through a retrainable set of layers.\\

\noindent The model was implemented using the PyTorch library, trained with a batch size of 32, L2 regularisation, and the AdamW optimiser (learning rate = $7.14\times10^{-4}$, momentum = $0.9$) \citep{Loshchilov2017}. The YOLOv8 classification configuration uses cross-entropy as the loss function and softmax activation for final predictions.

\subsection{Metrics used to evaluate performance}

\noindent Several metrics were used to evaluate the models' performance. These metrics include the overall accuracy of the model (Eq. \ref{e1}), the precision (Eq. \ref{e2}) and recall (Eq. \ref{e3}) calculated for each class, and finally the F1 score (Eq. \ref{e4}). These metrics were then used to evaluate and interpret the results of the predictions made by the model.\\

\noindent Using $\text{TP}_{i}$ as the number of true positive predictions made for class $i$, $\text{FN}_{i}$ as the number of false negative predictions made for class $i$, $\text{FN}_{i}$ as the number of false negative predictions made for class $i$, and N as the total number of test samples, the metrics were calculated respectively as:

\begin{equation}\label{e1}
    \text{ACC}_{\text{model}} = \frac{1}{\text{N}} {\sum_{i=1}^{3} \text{TP}_{i}} \times\text{100} , 
\end{equation}

\begin{equation}\label{e2}
    \text{Precision}_i = \frac{\text{TP}_i}{\text{TP}_i + \text{FP}_i},
\end{equation}

\begin{equation}\label{e3}
    \text{Recall}_i = \frac{\text{TP}_i}{\text{TP}_i + \text{FN}_i} ,
\end{equation}
\begin{center}
    \text{ and}
\end{center}
\begin{equation}\label{e4}
    \text{F1 score}_i = \frac{2\times \text{Precision}_i \times \text{Recall}_i}{\text{Precision}_i + \text{Recall}_i}\text{.}
\end{equation}

\section{Results and discussion}\label{Sec:results}

\noindent In this section, the training and testing results from the models that were trained will be discussed. From the results detailed in Table \ref{tab:SummaryofTrainingResultsTable}, it is clear that all the models performed better on the test set than on the validation set. Due to the sample size of the dataset, it can be attributed to the fact that random sampling techniques included fewer noisy samples in the test set. To further verify results, cross-validation techniques can be used, which are discussed as part of future work in Section \ref{Sec:conclusion}.\\

\begin{table}[h]
    \centering
    \begin{tabular}{c|cc}
         & Validation accuracy after training &  Test set accuracy \\
         \hline
        YOLOv8       & 89\%  & 92\% \\
        DenseNet-201 & 83\%  & 89\% \\
        VGGNet-19    & 81\%  & 88\% \\
        MobileNet    & 82\%  & 88\% \\
        CNN          & 80\%  & 87\% \\
        ResNet-152   & 80\%  & 87\% \\ 
    \end{tabular}
    \captionsetup{width=0.9\textwidth}
    \caption{Summarised training results for the models.}
    \label{tab:SummaryofTrainingResultsTable}
\end{table}

\noindent The results for each model were determined according to the already defined metrics, and a summary can be found in Table \ref{tab:SummaryofResultsTable}.\\

\begin{table}[h]
    \centering
    \begin{tabular}{c|ccc}
         & Average precision &  Average Recall & Average F1 score \\
         \hline
        YOLOv8          & 92\%  & 92\% & 92\%   \\
        DenseNet-201    & 89\%  & 89\% & 89\%   \\
        ResNet-152      & 89\%  & 89\% & 89\%   \\ 
        MobileNet       & 88\%  & 88\% & 88\%   \\ 
        CNN             & 87\%  & 87\% & 87\%   \\ 
        VGGNet-19       & 87\%  & 86\% & 87\%   \\ 
    \end{tabular}
    \captionsetup{width=0.9\textwidth}
    \caption{Performance indicators for the five models trained with the stratified dataset. All numbers are rounded to integer values.}
    \label{tab:SummaryofResultsTable}  
\end{table}

\noindent For the VGGnet-19 model, it achieved an overall classification accuracy of 88\%. Precision, recall, and F1 scores for the individual classes are consistently high, with the model demonstrating particularly strong performance in classifying the Empty class, achieving a precision of 91\%. However, the model's performance for the Type III class is slightly lower, with precision, recall, and F1 scores all in the low 80s.\\

\noindent The MobileNet model also achieved an overall accuracy of 88\%, comparable to that of VGGnet-19. This model achieved a high precision for the Empty class (93\%), but lower values for the Type III class (84\%). The recall for the Type II class is particularly strong at 96\%, but the performance for the Type III class is still relatively weaker, with a recall of 81\%.\\

\noindent ResNet-152, with an overall accuracy of 87\%, shows a higher recall for the Type II class (94\%) compared to its precision (91\%). However, its performance in the Empty and Type III classes is somewhat weaker, with precision and recall for Type III dropping to 79\% and 85\%, respectively. The model struggles to distinguish between these two classes.
\begin{figure}[h]
  \centering
    \includegraphics[width=.8\textwidth]{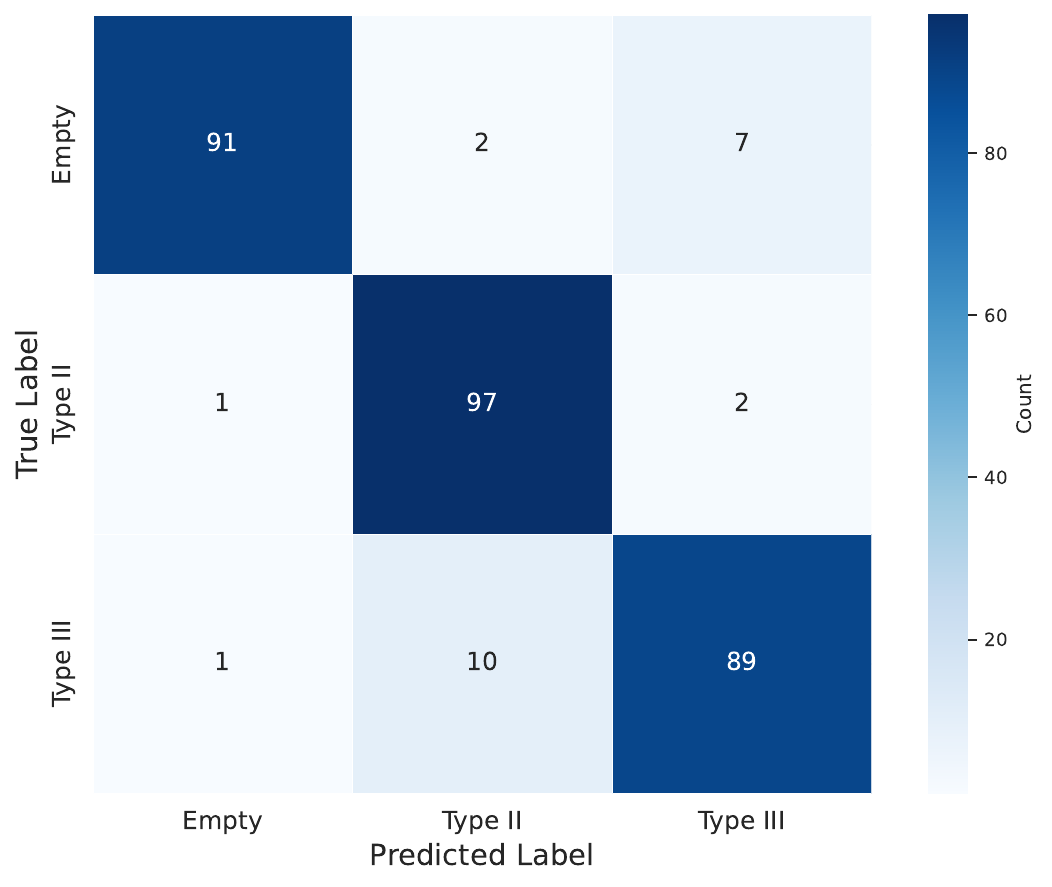}
    \captionsetup{width=0.9\textwidth}
    \caption{Confusion matrix illustrating the distribution of classifications made by the best performing (YOLOv8) model. Values represent raw counts of predictions for each class (Empty, Type II and Type III), with rows as true labels and columns as predicted labels made by the model.}
    \label{fig:Conf}
\end{figure}%
\noindent The DenseNet-201 model performed slightly better than ResNet-152, achieving an overall accuracy of 89\%. The metrics for this model indicate strong performance across the board, especially for the Empty and Type II classes. Precision, recall, and F1 scores for the Type III class are all in the mid-80s, indicating that DenseNet-201 performs well across all categories, with fewer misclassifications.\\

\noindent YOLOv8 outperformed all the other models, with an impressive overall accuracy of 92\%. The model exhibited excellent precision for the Empty class (98\%) and strong recall for the Type II class (98\%). Its F1 scores were consistently high across all classes, suggesting that YOLOv8 is the most balanced and reliable model in this experiment. There is, however, still room for improvement, with 89\% precision for the Type II class and 89\% recall for the Type III class.\\

\noindent When inspecting individual spectra that the model misclassified, it is clear that the model demonstrates some weaknesses. The model often mistakes Empty and Type III classes for Type II, likely due to noise or unlabeled artefacts. Some labelled Type II spectra exhibit features resembling Type III spectra, leading to errors. Additionally, Type III spectra with significant ``zebra'' noise are frequently misclassified as Type II, with the model showing reduced confidence in these predictions.\\

\noindent In summary, while all models demonstrated strong classification performance, YOLOv8 consistently outperformed the others, particularly in precision and recall, resulting in higher overall accuracy and more reliable classifications across classes.

\clearpage
\section{Conclusion}\label{Sec:conclusion}

In this paper, several pre-trained models were fine-tuned with transfer learning models for the purpose of classifying between spectrograms containing Type II SRBs, Type III SRBs, and Empty spectrograms.\\ 

\noindent The results of the transfer learning models show notable differences in performance, with YOLOv8 emerging as the most accurate and reliable, achieving an overall accuracy of 92\%. This model displayed strong precision and recall across all categories, especially for the Empty and Type II classes. In comparison, other models, such as VGGNet-19, MobileNet, ResNet-152, and DenseNet-201, showed decent overall performance but varied in their ability to classify certain categories, particularly Type III. Although these models achieved an overall accuracy of 87\%-89\%, their precision and recall for some classes, such as Type III, were less robust, with some misclassifications indicating areas for improvement.\\

\noindent Future work should focus on refining the models by incorporating more advanced data augmentation techniques and K-fold cross-validation to improve generalisation and the reliability of the results. These techniques can include existing image augmentation and the use of simulated radio bursts or generative ML models to create samples for underrepresented classes. Investigating ensemble methods and combining the strengths of different models may also provide a more reliable, balanced, and accurate solution to the task at hand. Future work should focus on automated parameter extraction after the identification process. This would allow large datasets to be created from archival data for large-scale statistical studies of the morphology of Type II and Type III SRBs.

\vspace{1em}
\noindent \bmhead{Supplementary information}
The code and the processed data used to train the models presented in this paper are available at \url{https://doi.org/10.5281/zenodo.15731453}.\\

\noindent \bmhead{Acknowledgements}
We thank FHNW in Brugg/Windisch, Switzerland, for providing data storage and process capacity. We also thank Australia-ASSA (Astronomical Society of South Australia), BIR (Rosse Observatory, Trinity College Dublin, Ireland) and EGYPT-Alexandria (Space Environment Research Lab, Institute of Basic and Applied Science, Egypt-Japan University of Science and Technology, New Borg El-Arab City, 21934 Alexandria, Egypt), as well as all the other observing stations contributing to the network. \\

\noindent \bmhead{Author Contributions}
The manuscript was written by Herman le Roux, incorporating significant input on methods and design, feedback, and revisions from G.R. Drevin, R.D. Strauss, P.J. Steyn, P.T. Gallagher, M. Daly, S.A. Maloney, and J. Scully. All authors participated in the manuscript review.\\

\noindent \bmhead{Data Availability} The raw data used in the study can be found in the {e-Callisto} data repository (\url{https://www.e-callisto.org/Data/data.html}). The processed data and code are available in the supplementary information files.\\

\section*{Declarations}

\noindent \bmhead{\text{Competing Interests}} The authors declare that there are no competing interests.\\

\noindent \bmhead{Open Access} This article is licensed under a Creative Commons Attribution 4.0 International License, which permits use, sharing, adaptation, distribution, and reproduction in any medium or format, as long as you give appropriate credit to the original author(s) and the source, provide a link to the Creative Commons licence,and indicate if changes were made. The images or other third party material in this article are included in the article’s Creative Commons licence, unless indicated otherwise in a credit line to the material. If material is not included in the article’s Creative Commons licence and your intended use is not permitted by statutory regulation or exceeds the permitted use, you will need to obtain permission directly from the copyright holder.
To view a copy of this licence, visit \href{http://creativecommons.org/licenses/by/4.0/}{http://creativecommons.org/licenses/by/4.0/}.

\clearpage

\backmatter

\clearpage
\bibliography{sn-bibliography}

\end{document}